\documentclass[final]{svjour3}
\usepackage{graphicx}
\usepackage{rotating}
\usepackage{amssymb}
\usepackage{mathptmx}
\usepackage{cite}
\usepackage{subfig}
\usepackage{diagbox}
\usepackage{xcolor}
\makeatletter
\journalname{Journal of Low Temperature Physics}

\begin{document}

\newcommand{\hdblarrow}{H\makebox[0.9ex][l]{$\downdownarrows$}-}
\title{Modelling TES non-linearity induced by a rotating HWP in a CMB polarimeter}

\author{T.~Ghigna\textsuperscript{{\normalfont \textit{a}, $ \dagger$}} \and T.~Matsumura\textsuperscript{{\normalfont \textit{a}}} \and Y.~Sakurai\textsuperscript{{\normalfont \textit{a}}} \and R.~Takaku\textsuperscript{{\normalfont \textit{b}}} \and K.~Komatsu\textsuperscript{{\normalfont \textit{c}}} \and S.~Sugiyama\textsuperscript{{\normalfont \textit{d}}} \and Y.~Hoshino\textsuperscript{{\normalfont \textit{d}}} \and N.~Katayama\textsuperscript{{\normalfont \textit{a}}}}

\institute{
\textsuperscript{\textit a}Kavli IPMU (WPI), UTIAS, The University of Tokyo, Kashiwa, Chiba 277-8583, Japan, 
\textsuperscript{\textit b}Department of Physics, University of Tokyo, 7-3-1,  Hongo, Bunkyo-ku, Tokyo 113-0033, Japan,
\textsuperscript{\textit c}Okayama University,  3-1-1, Tsushimanaka, Kita-ku, Okayama City, Okayama, 700-8530, Japan,
\textsuperscript{\textit d}Saitama University, 255 Shimookubo, Sakura-ku, Saitama, 338-8570, Japan.
\textsuperscript{$ \dagger$}Corresponding author: tommaso.ghigna@impu.jp.
}

\maketitle

\begin{abstract}
Most upcoming CMB experiments are planning to deploy between a few thousand and a few hundred thousand TES bolometers in order to drastically increase sensitivity and unveil the B-mode signal. Differential systematic effects and $1/f$ noise are two of the challenges that need to be overcome in order to achieve this result. In recent years, rotating Half-Wave Plates have become increasingly more popular as a solution to mitigate these effects, especially for those experiments that are targeting the largest angular scales. However, other effects may appear when a rotating HWP is being employed. In this paper we focus on HWP synchronous signals, which are due to intensity to polarization leakage induced by a rotating cryogenic multi-layer sapphire HWP employed as the first optical element of the telescope system. We use LiteBIRD LFT as a case study and we analyze the interaction between these spurious signals and TES bolometers, to determine whether this signal can contaminate the bolometer response. We present the results of simulations for a few different TES model assumptions and different spurious signal amplitudes. Modelling these effects is fundamental to find what leakage level can be tolerated and minimize non-linearity effects of the bolometer response.  
\keywords{CMB, Bolometers, Transition Edge Sensors, Half-Wave Plates}
\end{abstract}

\vspace{-6mm}
\section{Introduction}
\vspace{-3mm}
Probing the primordial CMB B-mode signal is one of the main objectives of modern cosmology. While still an open question, a very rapid inflationary expansion of the Universe is expected from current theories. The best available tool to probe inflation is a measurement of the polarized CMB signal \cite{zaldarriaga_seljak97, kamionkowski97}. On large angular scales ($\ell \lesssim 200$) the polarized B-mode signal is expected to be dominated by a primordial inflationary component (if the tensor-to-scalar ratio $r\gtrsim0.01$, or even $r\gtrsim0.001$ for $\ell < 10$)\footnote{Galactic foregrounds and gravitational lensing are two major challenges for a B-mode measurement. While a detailed discussion is beyond the scope of this paper, we briefly mention them here because their presence can make non-linear effects even more relevant.}. Two of the main challenges for CMB experiments are atmospheric fluctuations (for ground-based experiments) and receiver stability on long time-scales ($1/f$ noise and gain fluctuations). Therefore, several present and future CMB experiments (CLASS \cite{Dahal2021}, Simons Array \cite{Suzuki2016}, Simons Observatory \cite{simonsObs2019}, LiteBIRD \cite{Hazumi2020}) are employing (or planning to) polarization modulators, such as continuously-rotating Half-Wave Plates (HWP). A polarization modulator up-converts the polarized sky signal to a frequency range where the noise spectrum is expected to be purely white \cite{Kusaka2014}. In a nutshell, if the HWP rotates at a frequency $f_{hwp}$, the polarized signal is modulated at $f_{mod}=4f_{hwp}$, therefore by properly choosing $f_{hwp}$ it is possible to shift the signal of interest above the $1/f$ knee-frequency.
\vspace{0mm}

LiteBIRD is one of the experiments that will adopt this strategy. The LiteBIRD space mission consists of 3 telescopes
each equipped with a HWP, as the first optical element, and a cryogenic focal plane ($\sim 100$ mK) populated with polarization sensitive Transition-Edge Sensor (TES) bolometers. At Kavli IPMU we are in charge of developing the HWP and the rotation mechanism for LFT. The current baseline consists of a multi-layer sapphire HWP with anti-reflection layers laser-machined directly on the HWP surfaces \cite{Komatsu2019, Takaku2021}. Even though HWPs are effective in rejecting $1/f$ noise, as well as removing some systematic effects arising from intrinsic differences among detectors (e.g. bandpass and beam mismatch, gain mismatch, etc.), HWP non-idealities can introduce other systematic effects. For example, in \cite{Takakura2017, Didier2017} (POLARBEAR and EBEX collaborations) the authors identified large\footnote{Amplitudes of the order of hundreds of mK. For comparison the CMB solar dipole is $\sim\pm 3.362$ mK.} signals due to instrumental intensity-to-polarization leakage. In both cases the signal is attributed primarily to optical elements before the HWP, and it is modulated by the HWP at $4f_{hwp}$.\footnote{Any optical element located before the HWP generates instrumental polarization that is going to be modulated by the HWP at the same frequency of the sky signal. Having the HWP as the first optical element allows for a clear separation of instrumental and celestial polarization.} In both analysis the authors argue that this signal affects the TES response by driving it into a non-linear regime.

\vspace{0mm}
In the case of LiteBIRD and Simons Observatory the HWP is going to be the first optical element (if we neglect windows or thermal filters), therefore we should not expect the same $4f_{hwp}$ leakage level, however HWP non-idealities or non-orthogonal incidence can create similar effects.

In this paper we address the problem of TES non-linearity and highlight how this can couple to HWP synchronous signals (HWPSS). We use LiteBIRD LFT as a case study. First, we define TES non-linearity and HWP intensity-to-polarization leakage.
\begin{figure}[htbp]
    \vspace{-5mm}
    \centering
    \subfloat{{\includegraphics[width=.45\textwidth]{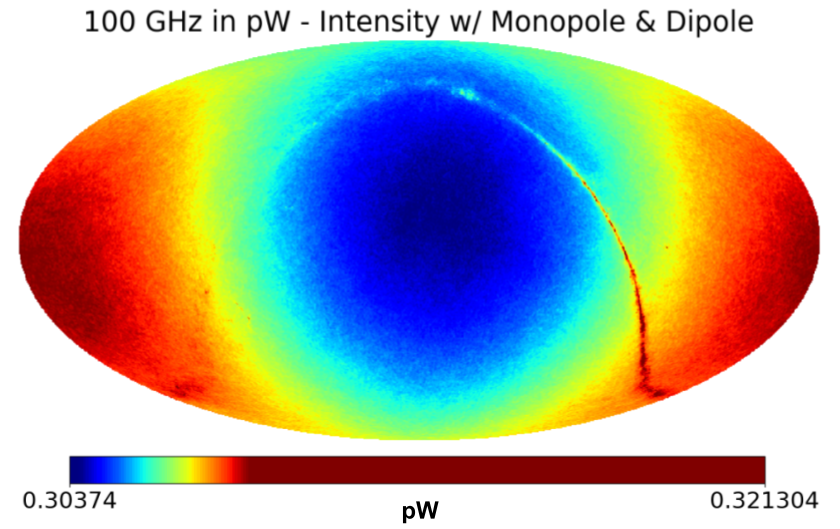} }}%
    \subfloat{{\includegraphics[width=.44\textwidth]{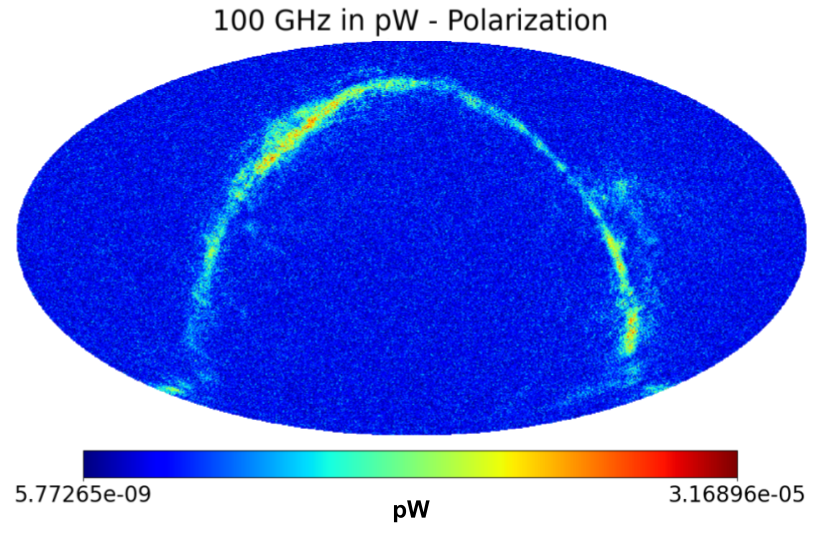} }}%
    \vspace{-3mm}
    \caption{Example of the input maps in pW at 100 GHz. The left plot shows the total intensity map including CMB anisotropies, foreground (generated with \textit{pysm}) the CMB monopole and dipole and the telescope loading for a LiteBIRD LFT 100 GHz detector according to \cite{Westbrook2021, Hasebe2021}. The right plot shows the polarization map $\sqrt{Q^2+U^2}$. The maps are generated in CMB units and then converted to input power. (Color figure online.)}%
    \label{fig:simulation_map}
\end{figure}
\begin{figure}[htbp]
    \vspace{-5mm}
    \centering
    \subfloat{{\includegraphics[width=.45\textwidth]{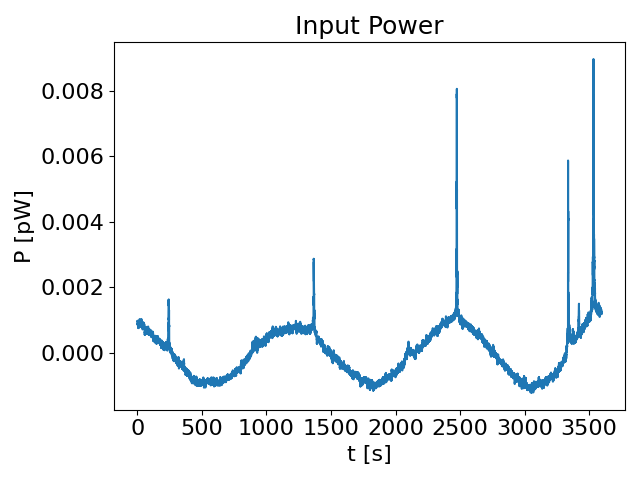} }}%
    \subfloat{{\includegraphics[width=.45\textwidth]{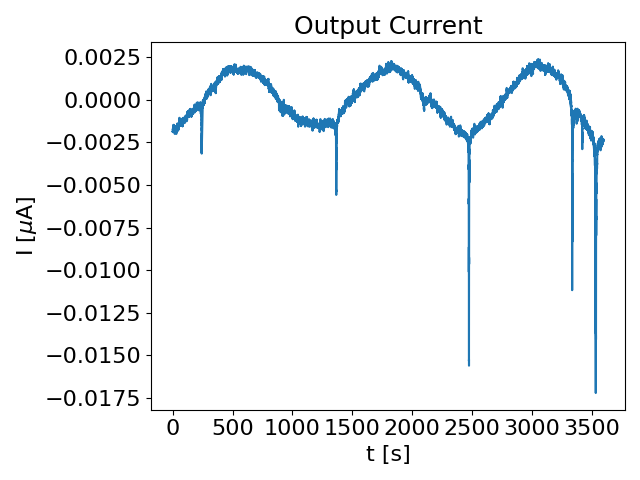} }}%
    \vspace{-4mm}
    \caption{Left: 1 hour of noiseless time-ordered data without IP leakage (ideal case) generated from the map in Figure \ref{fig:simulation_map} according to LiteBIRD scan strategy \cite{litebird_ptep} for a detector at the telescope boresight. Right: time-ordered data after processing the data on the left with the TES detector simulator. The data contains both total intensity and polarization. The presence of the dipole and galactic plane can be easily seen in the time-stream. (Color figure online.)}%
    \label{fig:simulation_tod}
    \vspace{-3mm}
\end{figure}
We present a code developed in \textit{python} to realistically simulate the detector response to the sky signal. Finally, we present preliminary results showing the impact of spurious signals (such as HWPSS) in the detector response. We compare these results to non-linearity estimates from POLARBEAR data \cite{Takakura2017}. We believe that the tool presented can be useful to forecast future experiment capabilities, define instrument requirements, inform the instrument design and develop methodologies to analyze the data and mitigate systematic effects. This is vital for the level of accuracy required by future experiments in particular for a space mission like LiteBIRD, which is aiming to achieve an accuracy in terms of tensor-to-scalar ratio of $\sigma_{r}\lesssim0.001$. A detailed study of LiteBIRD requirements can be found in \cite{litebird_ptep}.
\vspace{-6mm}
\section{TES and HWP model}\label{sec:model}
\vspace{-3mm}
\paragraph{Transition-Edge Sensor:} According to \cite{Irwin2005} we can define the behaviour of a DC voltage-biased bolometer by solving the system of coupled differential equations: 
\begin{eqnarray}
    L\frac{dI}{dt} & = & V_{b} - IR_{tes} - IR_{sh} \label{eq:diff_equations_I} \\
    C\frac{dT}{dt} & = & - P_{b} + I^2R_{tes} + P_{opt}.
    \label{eq:diff_equations_T}
\end{eqnarray}
It is commonly understood that, in the small signal ($\delta P_{opt} \ll \bar{P}_{opt}$) and high loop-gain limits $\mathcal{L}\ll 1$, the detector current responsivity reduces to $S_{I}\sim -1/V_{b}$ (for a "slow" signal: $\omega \rightarrow 0$).

In \cite{Takakura2017, TakakuraPHD} the authors present a non-linearity model where the detector responsivity is loosely dependent from the input power variations $\delta P_{opt}$. This takes the following form for the detector current:
\begin{equation}
    \delta I(t) = [S_I + S^{\prime}_I \delta P_{opt}(t)]\delta P_{opt}(t - \tau^{\prime}\delta P_{opt}(t)),
        \label{eq:current_responsivity_nonlinear}
\end{equation}
where $S^{\prime}_I$ and $\tau^{\prime}$ are non-linear terms. In the following, after briefly describing the HWP, we will discuss how the non-linear term couples to the HWP creating spurious signals.
\vspace{-3mm}
\paragraph{Half-Wave Plate:} The effect of a rotating HWP on the incoming sky signal is defined in terms of its Mueller matrix $\Gamma_{hwp}$ \cite{Komatsu2019}. For an ideal HWP $\Gamma_{hwp}$ is a diagonal matrix of the form $diag(1,1,-1,-1)$.
However, for a real HWP the off-diagonal elements of the matrix $\Gamma_{hwp}$ can be non-zero, and cause mixing between the $I$, $Q$ and $U$ components of the input signal. In particular, in the context of this paper, 
\begin{figure}[htbp]
    \vspace{-5mm}
    \centering
    \subfloat{{\includegraphics[width=.4\textwidth]{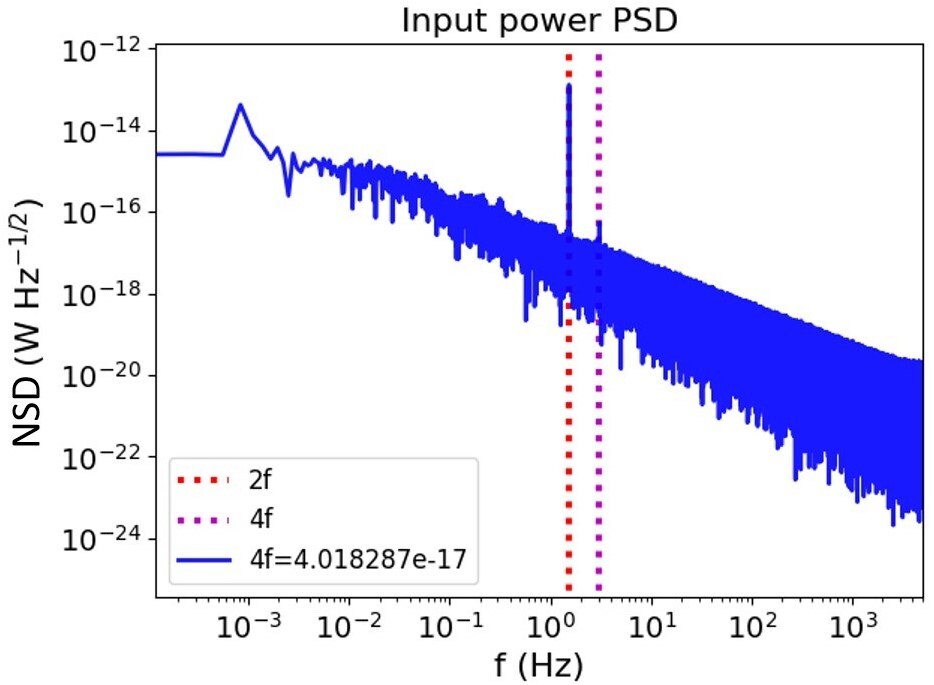} }}%
    \subfloat{{\includegraphics[width=.4\textwidth]{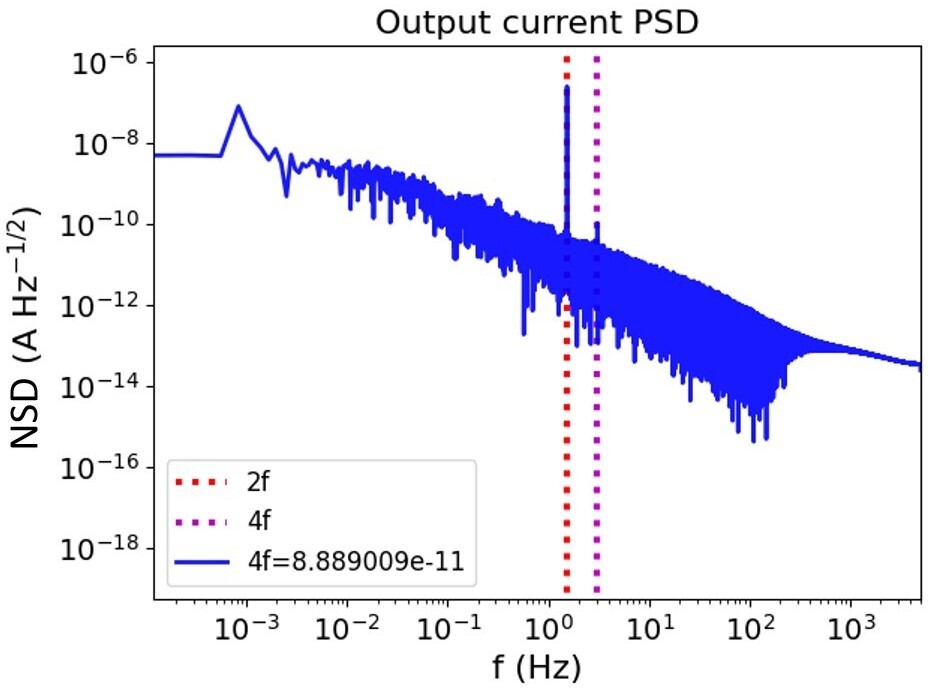} }}%
    \vspace{-3mm}
    \subfloat{{\includegraphics[width=.4\textwidth]{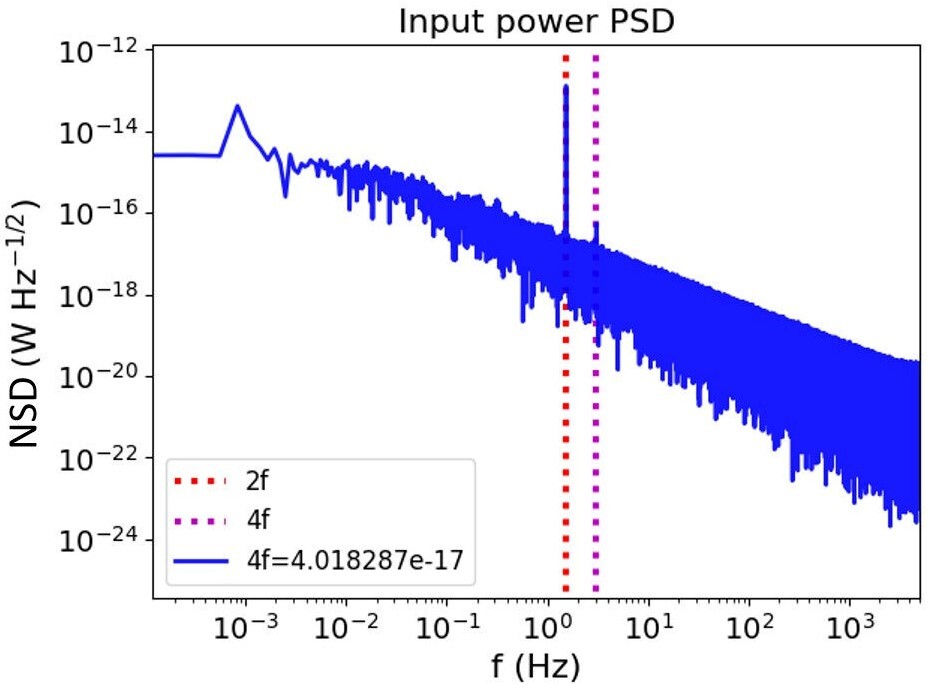} }}%
    \subfloat{{\includegraphics[width=.4\textwidth]{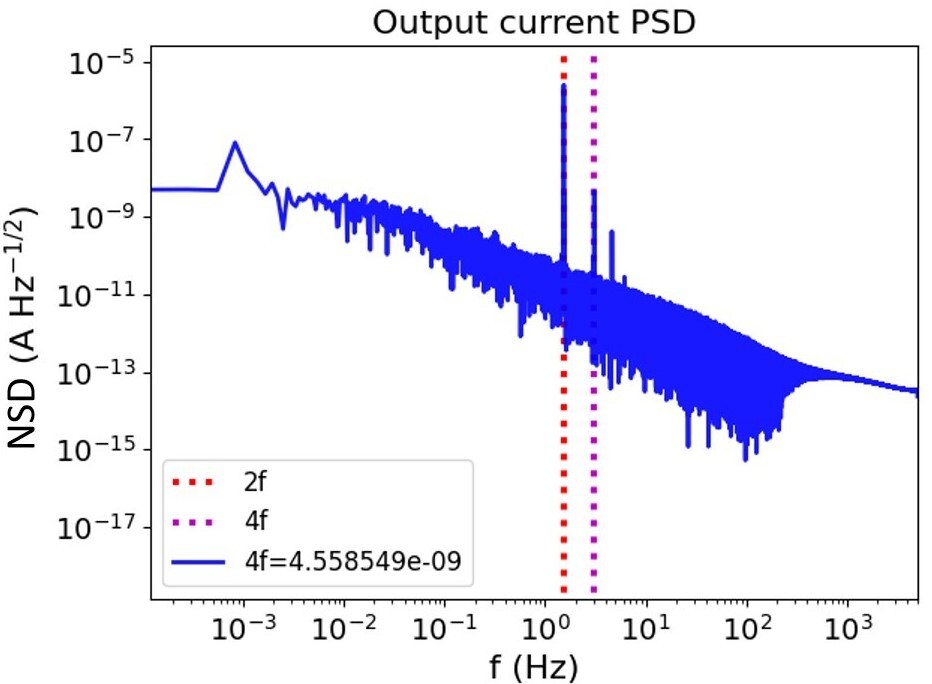} }}%
    \vspace{-3mm}
    \caption{Top: 1\% IP leakage. Left: NSD (noise spectral density) of the input TOD in units of power (1 hour). Right: NSD of the TOD after processing with the TES simulator. Bottom: Same as top for 10\% IP leakage. These simulations are noiseless, therefore the spectrum is dominated by the foreground signal. (Color figure online.)}%
    \label{fig:simulation_fft}
    \vspace{-3mm}
\end{figure}
we are interested in the $\gamma_{QI}$ and $\gamma_{UI}$ components that can cause intensity-to-polarization leakage (IP). From \cite{Komatsu2019} we find that non-zero values can result in a $2 f_{hwp}$ signal of the form: $\frac{1}{2}\sqrt{\gamma^2_{QI}+\gamma^2_{UI}}I_{in}\cos(2\omega_{hwp} t-2\arctan{\frac{\gamma_{UI}}{\gamma_{QI}}})$ \cite{Takaku2021}. Other effects can also appear, if we take into consideration the position of the HWP in the optical system, but for the scope of this paper we limit the discussion to the HWP intensity-to-polarization leakage. However, this analysis can be easily generalized to other cases.
\vspace{-7mm}
\section{Simulation}\label{sec:sims}
\vspace{-3mm}
As discussed in Section \ref{sec:model}, if the TES responsivity depends loosely on the incoming signal $\delta P_{opt}$ the bolometer response becomes non-linear. From Equation \ref{eq:current_responsivity_nonlinear} we can easily see that a signal at a generic frequency $\omega$ can partially leak to $2\omega$.\footnote{From simple trigonometry, if $\delta P_{opt}\sim\sin{\omega t}$, then the term $(\delta P_{opt})^2\sim\frac{1-\cos{2\omega t}}{2}$.} Hence, the $2 f_{hwp}$ IP signal discussed in Section \ref{sec:model} can partially leak to $4 f_{hwp}$, where the polarized signal is expected.
In order to perform these simulations and address the impact of non-linearity on cosmological data we have developed a $python$ module\footnote{https://github.com/tomma90/tessimdc} that can be plugged into a more standard time-order data (TOD) simulator. First, the input sky maps are converted to units of power by integrating the sky emission ($Jy/sr$) over the beam function $B(\Omega, \nu)$ and the band-pass function $G(\nu)$. An example is shown in Figure \ref{fig:simulation_map} for LiteBIRD LFT 100 GHz. We use the $pysm$\footnote{https://pysm3.readthedocs.io/en/latest/} package to generate the sky model. We then add the CMB dipole and the expected optical loading (for a LiteBIRD LFT 100 GHz detector) due to the CMB monopole and telescope loading: $\bar{P}_{opt}=0.3061$ pW \cite{Westbrook2021, Hasebe2021}. 

The second step consists in generating the telescope pointing according to LiteBIRD scan strategy \cite{Hazumi2020} which is used to generate the TOD in units of power. At this point in the simulation, we can add a HWP synchronous signal. 
\begin{figure}[htbp]
    \centering
    \includegraphics[width=1\textwidth]{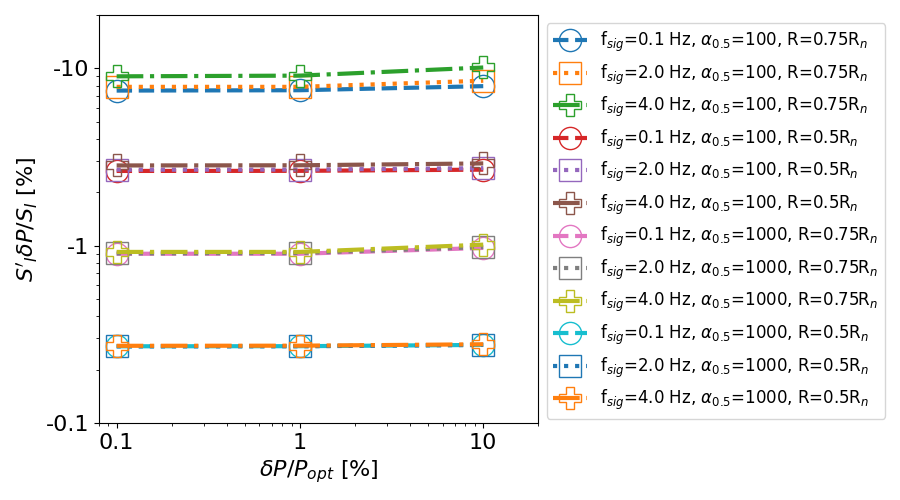}
    \vspace{-8mm}
    \caption{TES detector non-linearity. We simulated representative sinusoidal signals of frequencies $f_{sig}=$ 0.1, 2 and 4 Hz and amplitudes between 0.1\% and 10\% of the expected average optical loading $P_{opt}$ to test non-linearity according to the formula in Equation \ref{eq:current_responsivity_nonlinear}. We tested 2 cases: a TES with log-responsivity $\alpha_{0.5\Omega}$=100 and $\alpha_{0.5\Omega}$=1000 at $R=0.5\times R_{n}$. For each case we biased the TES at $R=0.5\times R_{n}$ and $R=0.75\times R_{n}$ and studied its response. It is clear that a narrower transition and a higher loop-gain lead to a more linear device. (Color figure online.)}%
    \label{fig:simulation_nonlin}
    \vspace{-5mm}
\end{figure}
At present we can only fix the amplitude of the signal. We are planning to implement the possibility to input a measured or simulated HWP Mueller matrix to allow for more realism to the simulation. 
The TOD is then fed to the TES simulator as the input optical power $P_{opt}$ in Equation \ref{eq:diff_equations_T}. An example for a 1-hour scan is shown in the left plot of Figure \ref{fig:simulation_tod}. 
\vspace{0mm}

The TES response simulator consists of a module based on the Runge-Kutta 4th order method to solve Equations \ref{eq:diff_equations_I} and \ref{eq:diff_equations_T}. We have made some assumptions to set the parameters according to LiteBIRD specifications. The thermal conductance $G$ and heat capacity $C$ are defined assuming a normal time-constant $\tau_{0}=C/G\sim30$ ms and a saturation power $P_{sat}=2.5\times \bar{P}_{opt}$ (average optical power defined by the CMB monopole and telescope loading).
\vspace{0mm}

Another key assumption is the TES resistance. In the simulations carried out for this paper, we have assumed a simple $arctan$ model where the resistance depends only on the TES temperature. We defined the normal resistance $R_n = 1 \Omega$ and the log-responsivity ($\alpha=\frac{d log R}{d log T}$) $\alpha_{0.5\Omega}=100$ at $0.5\times R_n$ (conservative value). In the future we will explore other models that do not neglect the dependence of the resistance from the TES current (two-fluids, RSJ, etc.)
One shortcoming of this procedure is the need for a high sampling rate compared to more traditional TOD simulator. This is due to the need to solve iteratively the differential equations at a rate faster than the detector effective time-constant ($\tau_{eff}\sim\tau_{0}/(\mathcal{L}+1)$) in order to avoid numerical errors. For high-loop gain conditions ($\mathcal{L}\sim 10$), this translates to a sampling rate $\gtrsim$ 1 kHz. 

In the right plot of Figure \ref{fig:simulation_tod} we can see the result of the TES simulator for an ideal case without HWPSS. In Figure \ref{fig:simulation_fft} we show the impact of a $2 f_{hwp}$ ($\omega_{hwp} = 46$ rpm) HWPSS component of $1\%$ and $10\%$ amplitudes (relative to the expected optical loading $\bar{P}_{opt}$)\footnote{The data used in this work are noiseless to focus on the detector response characterization.}. A larger $2 f_{hwp}$ signal clearly results into an enhancement of the $4f_{hwp}$ component in the NSD (noise spectral density), which is due to the non-linear term of Equation \ref{eq:current_responsivity_nonlinear} that grows with the amplitude of the signal ($S^{\prime}_{I}\delta P^2$).

Finally, we test the non-linearity model in Equation \ref{eq:current_responsivity_nonlinear} using representative signals (for this simulation we simplified the input compared to Figure \ref{fig:simulation_fft}, by including only the monopole - $P_{opt}$ - and a signal of frequency $f_{sig}$ and relative amplitude $\delta P/P_{opt}$) of frequency $f_{sig}$ between 0.1 Hz and 4 Hz and amplitudes between 0.1\% and 10\% of the expected optical loading for two transition cases: a log-responsivity $\alpha_{0.5\Omega}$=100 and $\alpha_{0.5\Omega}$=1000. For all cases we study the response when the detector is biased at $R=0.5\times R_{n}$ and $R=0.75\times R_{n}$. After processing the input signal with the TES simulator we convert the TES current $\delta I$ to input power $\delta P^{\prime}=\delta I/S_{I}$, in the high loop-gain limit. Afterwards we compare the input signal $\delta P$ and the calibrated output $\delta P^{\prime}$ to determine the non-linear term $S^{\prime}_{I}\delta P$ for each case: $S^{\prime}_{I}\delta P/S_{I}=(\delta P^{\prime}-\delta P)/\delta P$. 
In Figure \ref{fig:simulation_nonlin} we show a summary of the results. By isolating the term $S^{\prime}_{I}\delta P$ we find that this appears to depend on the TES model and bias conditions. In fact, the cases with steeper and narrower transition ($\alpha_{0.5\Omega}$=1000) and lower operating resistance (higher loop-gain) show reduced non-linearity levels.
\vspace{-6mm}
\section{Conclusions}
\vspace{-3mm}
We believe that the tool developed and presented in this paper is going to be very valuable going forward with the definition of the mission, both in terms of informing the instrument development and the data analysis. This is particularly true for the requirements of a space mission. We have given a preliminary assessment of the TES non-linearity model developed and tested on POLARBEAR data in \cite{TakakuraPHD, Takakura2017}. We found indications of the input power-dependent non-linear component. While the responsivity term seems to mimic well the model, we need a more in depth analysis of the non-linearity model to address the time-constant term. A more thorough analysis with propagation to map-making and cosmological analysis is needed to understand the full impact of this effect on the scientific output (e.g. tensor-to-scalar ratio). The results\footnote{The data that support the findings of this study are available from the corresponding author upon reasonable request.} presented in this paper are a starting point for a more in-depth analysis. More data and laboratory tests are needed to define realistic detector and instrument models for LiteBIRD. 

However, we can notice that in \cite{Takakura2017} the level of non-linearity of POLARBEAR data has been estimated to be in the range 0.3-0.8\% (varying among detector wafers). This is in agreement with the values found in our analysis for the cases with $\alpha_{0.5\Omega}=1000$ in Figure \ref{fig:simulation_nonlin}, where we have determined $S^{\prime}_I\delta P/S_I$ to be in the range 0.2-0.9\% (depending on the bias conditions). In our analysis we have also studied the case for $\alpha_{0.5\Omega}=100$ which is regarded as a pessimistic scenario, and in fact leads to larger non-linearity levels (as high as $\sim 9$\% according to Figure \ref{fig:simulation_nonlin}). This result clearly shows that a narrower superconductive transition improves the linearity of the device.

In conclusion, the absence of a turbulent atmosphere that affects ground-based experiments will allow LiteBIRD (or any other space mission) to achieve an unprecedented sensitivity. We can foresee that more stable observation conditions will reduce the impact of $1/f$ noise and gain fluctuations due to long time-scale variations of the optical loading. Hence, more subtle non-linear effects could become more dominant, which will require more realistic models of the detectors and read-out systems to forecast and mitigate them.
\vspace{-3mm}
\begin{acknowledgements}
We thank all LiteBIRD collaborators for support and help. In particular Juan Mac\'{\i}as-P\'erez and Satoru Takakura for usuful comments and feedback on the manuscript. This work was supported by JSPS KAKENHI Grant Numbers 22K14054, 18KK0083 and 19K14732. Kavli IPMU is supported by World Premier International Research Center Initiative (WPI), MEXT, Japan.
\end{acknowledgements}

\end{document}